\documentclass[twocolumn, times, tighten,twocolappendix]{aastex631}

\usepackage{graphicx,multirow,hyperref,url,color,xspace}
\usepackage{amsmath,amssymb}

\newcommand{\msun}{{\rm M}_{\sun}}
\newcommand{\rsun}{{\rm R}_{\sun}}
\newcommand{\lsun}{{\rm L}_{\sun}}

\newcommand{\xmm}{{\textit{XMM-Newton}}\xspace}

\newcommand{\swift}{{\textit{Neil Gehrels Swift}}\xspace}
\newcommand{\gaia}{{\textit{Gaia}}\xspace}

\newcommand{\source}{{MAXI J1820+070}\xspace}

\newcommand{\appropto}{\mathrel{\vcenter{
  \offinterlineskip\halign{\hfil$##$\cr
    \propto\cr\noalign{\kern2pt}\sim\cr\noalign{\kern-2pt}}}}}

\begin{document}

\title{The Donor of the Black-Hole X-Ray Binary MAXI J1820+070}
\shorttitle{The Donor of MAXI J1820+070}

\author[0000-0003-3457-0020]{Joanna Miko{\l}ajewska}
\affiliation{Nicolaus Copernicus Astronomical Center, Polish Academy of Sciences, Bartycka 18, PL-00-716 Warszawa, Poland}
\author[0000-0002-0333-2452]{Andrzej A. Zdziarski}
\affiliation{Nicolaus Copernicus Astronomical Center, Polish Academy of Sciences, Bartycka 18, PL-00-716 Warszawa, Poland}
\author{Janusz Zi{\'o}{\l}kowski}
\affiliation{Nicolaus Copernicus Astronomical Center, Polish Academy of Sciences, Bartycka 18, PL-00-716 Warszawa, Poland}
\author[0000-0002-5297-2683]{Manuel A. P. Torres}
\affiliation{Instituto de Astrof{\'{\i}}sica de Canarias, E-38205 La Laguna, Tenerife, Spain}
\affiliation{Departamento de Astrof{\'{\i}}sica, Universidad de La Laguna, E-38206  La Laguna, Tenerife, Spain}
\author[0000-0001-5031-0128]{Jorge Casares}
\affiliation{Instituto de Astrof{\'{\i}}sica de Canarias, E-38205 La Laguna, Tenerife, Spain}
\affiliation{Departamento de Astrof{\'{\i}}sica, Universidad de La Laguna, E-38206  La Laguna, Tenerife, Spain}

\shortauthors{Miko{\l}ajewska et al.}

\begin{abstract}
We estimate the parameters of the donor of the accreting black-hole binary MAXI J1820+070. The measured values of the binary period, rotational and radial velocities and constraints on the orbital inclination imply the donor is a subgiant with the mass of $M_2\approx 0.49^{+0.10}_{-0.10}\msun$ and the radius of $R_2\approx 1.19^{+0.08}_{-0.08}\rsun$. We re-analyze the previously obtained optical spectrum from the Gran Telescopio Canarias, and found it yields a strict lower limit on the effective temperature of $T>4200$ K. We compile optical and infrared fluxes observed during the quiescence of this system. From the minima $r$ and $i$-band fluxes found in Pan-STARSS1 Data Release 2 pre-discovery imaging and for a distance of $D\approx3$\,kpc, reddening of $E(B$--$V)=0.23$ and $R_2\approx{1.11R_\odot}$, we find $T\lesssim4230$\,K, very close to the above lower limit. For a larger distance, the temperature can be higher, up to about 4500\,K (corresponding to a K5 spectral type, preferred by previous studies) at $D=3.5$\,kpc, allowed by the Gaia parallax. We perform evolutionary calculations for the binary system and compare them to the observational constraints. Our model fitting the above temperature and radius constraints at $D\approx 3$\,kpc has the mass of $0.4M_\odot$, $T\approx4200$ K and solar metallicity. Two alternative models require $D\gtrsim 3.3$--3.4 kpc at $0.4 M_\odot$, $T\approx4500$ K and half solar metallicity, and $0.5M_\odot$, $T\approx4300$ K and solar metallicity. These models yield mass transfer rates of $\sim\!\!10^{-10}M_\odot$/yr, compatible with those based on the estimated accreted mass of $\approx\!2\times 10^{25}$\,g and the time between the 2018 discovery and the 1934 historical outburst.
\end{abstract}

\section{Introduction}
\label{intro}

The outburst of the transient accreting low-mass X-ray binary (LMXB) \source was discovered on 2018-03-06 in the optical range with the $V$-band magnitude of 14.88 \citep{Tucker18}. It was detected in X-rays six days later (on 2018-03-11; \citealt{Kawamuro18}). Given the high observed brightness and long duration of the outburst, it was the subject of a large number of observing campaigns, whose results have led to numerous insights into the nature of this source. \citet{Torres19,Torres20} have unambiguously determined that the accretor is a black hole (BH). However, the donor parameters have remained relatively loosely constrained (see Sections \ref{binary}--\ref{metal}). More accurate determinations of the donor radius and temperature are important for modelling the emission of the accretion flow in the quiescent state of this system \citep{Poutanen22}. Also, the metallicity of \source is of importance for calculation of the inner radius of the accretion disk during the outburst. On one hand, \citet{Buisson19} found the Fe abundance to be higher than solar by a factor of $\approx$3--9 and the disk extending close to the innermost stable circular orbit from fitting reflection model spectra to X-ray data in the luminous hard state. On the other, fits by \citet{Zdziarski21b} obtained a fractional Fe abundance of $\approx$1.1--1.6 and a truncated disk. However, Fe abundances from spectral fits are subject to significant systematic errors related to the effects of the assumed density of the reflector \citep{Garcia18n}. Therefore, an independent estimate of the abundances of the donor would be highly valuable in constraining the results of X-ray spectroscopy.

In this work, we consider evolutionary models of the binary, with the goal of constraining better the mass and radius of the stellar companion as well as its spectral type and metallicity. In Section \ref{binary}, we review available constraints on the distance and binary parameters, in Section \ref{photo}, we compile available mid-infrared to ultraviolet photometric measurements, which constrain the radius and temperature of the donor, in Section \ref{metal}, we consider its metallicity, in Section \ref{fluence}, we estimate the average mass accretion rate, and in Section \ref{evolution}, we perform the evolutionary calculations. Section \ref{summary} provides a summary of our results.

\section{System parameters}
\label{parameters}

\begin{figure}
\centerline{\includegraphics[width=\columnwidth]{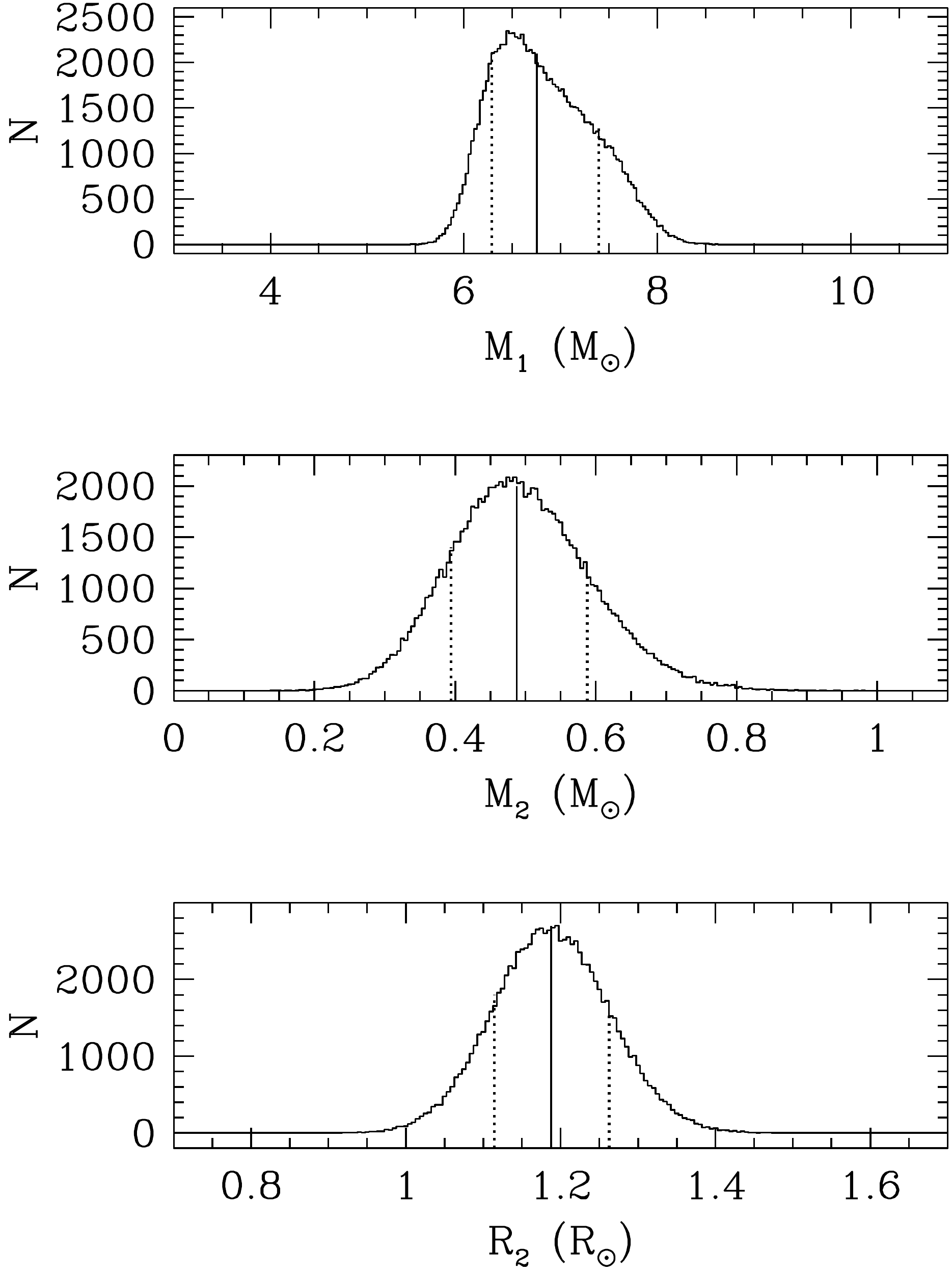}}
\caption{The results of Monte Carlo simulations (sampled $10^5$ times each) of the probability distribution for (top) the mass of the BH, (middle) the donor mass, and (bottom) the donor radius. The vertical solid lines mark the median, and the dashed lines show the 68\% confidence, ranges.
}
 \label{MC}
 \end{figure}

\subsection{The distance and binary parameters}
\label{binary}

\source is relatively nearby, with a distance of $D\approx 2.96\pm 0.33$\,kpc measured based on a radio parallax \citep{Atri20}. This agrees well with the current \gaia EDR3 parallax measurement of $D= 2.81^{+0.70}_{-0.39}$\,kpc \citep{Bailer_Jones21}. Then, \citet{Wood21} determined $D\leq 3.11\pm 0.06$\,kpc based on the proper motion of two moving transient ejecta \citep{MR94} during the hard-to-soft state transition, assuming the approaching and receding ejecta were identical. The upper limit of $3.17$\,kpc corresponds, however, to the minima of both angular velocities and to the bulk Lorentz factor of $\Gamma\rightarrow\infty$. The Lorentz factors of the transient ejecta measured in X-ray binaries are generally low, at most a few \citep{Miller-Jones06}. In the present case, $D=3$ and 3.1\,kpc correspond to rather large $\Gamma\geq 3.4$ and 5.1, respectively. Therefore, the measurement by \citet{Wood21} appears to imply $D\lesssim 3$\,kpc for realistic ejecta Lorentz factors. 

The binary orbital period is $P= 0.68549 \pm 0.00001$\,d \citep{Torres19} and the inclination of the binary is constrained at $3\sigma$ to $i\approx 66.2\degr$--$80.8\degr$ \citep{Torres20}. While the latter is consistent with the inclination of the jet, $i_{\rm j}\approx 64\pm 5\degr$ \citep{Atri20, Wood21}, the jet and binary axes have been estimated to be strongly misaligned \citep{Poutanen22}, in which case the jet inclination cannot be used to constrain the system masses.

\citet{Torres20} established the rotational line broadening of the companion of $v_{\rm rot}\sin i= 84\pm 5$\,km\,s$^{-1}$, which in combination of the radial velocity semi-amplitude ($K_2=417.7 \pm 3.9$\,km\,s$^{-1}$) yields (assuming cororation and Roche-lobe filling) the mass ratio of $q=M_2/M_1\approx 0.072\pm 0.012$, where $M_1$ and $M_2$ are the masses of the BH and its companion star. These uncertainties are 1-$\sigma$. \citet{Torres20}, following \citet{Marsh94}, obtained 2-$\sigma$ limits of $5.73 <M_1/\msun < 8.34$ and $0.28 < M_2/\msun <0.77$ by combining the 1-$\sigma$ uncertainties in the measured orbital parameters with the extreme limits for the range of allowed inclinations. On the other hand, here we have performed Monte Carlo simulations assuming Gaussian distributions of $v_{\rm rot}\sin i$ and $K_2$ (which give $q$ via eq.\ 4 of \citealt{zz17}), $P$, and a flat distribution of $\cos i$ within the range given above. This gives us probablilistic estimates of $M_1$ and $M_2$ and the donor radius, $R_2$ [$=P(v_{\rm rot}\sin i)/(2\pi \sin i)$]. The results are shown in Figure \ref{MC}, and they imply $M_2= 0.49^{+0.10}_{-0.10}\msun$, $R_2= 1.19^{+0.08}_{-0.08}\rsun$ and $M_1= 6.75^{+0.64}_{-0.46}\msun$, where we give the median and the 68\% confidence ranges.  Since the radii of main-sequence stars can be roughly approximated by $R_{\rm MS}/\rsun\approx (M/\msun)^{0.9}$, the above range of the donor radius implies that the donor is a moderately evolved star (a subgiant). 

The donor average spectrum from observations by the Gran Telescopio Canarias (GTC) during the quiescence after the outburst was found to approximately match those of K3--5 V main-sequence templates \citep{Torres20}. The effective temperatures of those stars are given by \citet{Pecaut13} as 4840--4450\,K. We have, however, re-analyzed that spectrum and found a strict limit of $T> 4200$\,K based on the lack of molecular bands (TiO and CN)  expected at $T\leq 4200$\,K. This limit approximately corresponds to a K6 stellar type, and we use it in our analyses below.

\begin{table*}\centering
\caption{The minimum observed fluxes for \source}
\begin{tabular}{ccccccc}
\hline
Source & Band & $\langle\lambda\rangle\,[\mu$m] & $F_{\nu,{\rm obs}}\,[\mu$Jy] & $\langle F_{\nu,{\rm obs}}\rangle\,[\mu$Jy] & $10^{A_\lambda/2.5}$&  Dates \\
\hline
PS1 & $y$  & 0.963& $265\pm 10$   &$290\pm 3$& 1.33 &2011-09-11, 2012-03-30\\
PS1 & $z$  & 0.868& $238\pm 1$    &$256\pm 1$& 1.40 &2013-04-24\\
PS1 & $i$  & 0.755& $141\pm 6$    &$220\pm 1$& 1.55 &2013-05-31\\
PS1 & $r$  & 0.622& $89.8\pm 4.5$ &$197\pm 1$& 1.77 &2013-05-31, 2013-06-19\\
PS1 & $g$  & 0.487& $54.9\pm 3.9$ &$117\pm 1$& 2.15 &2012-06-11, 2013-06-10\\
\hline
ZTF & $r$ & 0.622& $139\pm 5$ && 1.77& 2021-02-09\\
ZTF & $g$ & 0.487& $60.4\pm 4.1$ && 2.15& 2020-10-13\\
\hline
\end{tabular}
\tablecomments{Here and in Tables \ref{log} and \ref{Juri}, $F_{\nu,{\rm obs}}$ is the observed flux and the intrinsic one is $F_{\nu,{\rm obs}} 10^{A_\lambda/2.5}$ for $E(B$--$V)=0.23$ ($1\,{\rm Jy}\equiv 10^{-23}$\,erg\,cm$^{-2}$\,s$^{-1}$\,Hz$^{-1}$). For PS1 , the values of $F_{\nu,{\rm obs}}$ are averages from pairs of observations separated by $\lesssim$0.5\,h with the individual fluxes differing by $<$25\%, the uncertainties are the errors of those averages. The average fluxes from the PS1 stacked images, $\langle F_{\nu,{\rm obs}}\rangle$, are given for comparison. For ZTF, the flux uncertainties correspond to the original magnitude uncertainties. }
\label{pan}
\end{table*}

\setlength{\tabcolsep}{3pt}\begin{table*}\centering
\caption{Selected infrared and optical observations of \source in quiescent states}
\vskip -0.4cm                               
\begin{tabular}{lccccccll}
\hline
Source & Band & $\langle\lambda\rangle\,[\mu$m] & $F_{0}$\,[Jy]  &  $m_\lambda$ & $F_{\nu,{\rm obs}}\,[\mu$Jy] & $10^{A_\lambda/2.5}$&  Date  & Reference \\
\hline
WISE &$W2$       & 4.6  & 3631 & $17.95\pm 0.06$ &$239_{-13}^{+14}$ &1.02  &2010-01-14--2010-07-17& \citet{Cutri13}\\
WISE &$W1$       & 3.35 & 3631 & $17.59\pm 0.04$ &$334_{-11}^{+12}$ &1.04  &2010-01-14--2010-07-17& \citet{Cutri13}\\
2MASS &$K$      & 2.22 & 655   & $15.12\pm 0.12$ &$588_{-61}^{+69}$ &1.08 & 1999-07-23& \citet{Cutri03}\\
2MASS &$H$      & 1.60 & 1150  & $15.44\pm 0.10$ &$764_{-69}^{+75}$ &1.13 & 1999-07-23& \citet{Cutri03}\\
2MASS &$J$      & 1.26 & 1580  & $15.87\pm 0.07$ &$711_{-42}^{+45}$ &1.20   & 1999-07-23& \citet{Cutri03}\\
GSC2.3.2& $I$      & 0.812 & 2681  & $17.24\pm 0.43$ &$341_{-111}^{+166}$ & 1.46 & 1989--2000 & \citet{Lasker08}\\
USNO-B1.0   & $R$  & 0.658& 3247  & $16.76\pm 0.30$& $642_{-155}^{+204}$& 1.71 & 1949--1965 & \citet{Monet03}\\
GSC2.3.2 & $R$  & 0.658& 3247  & $18.05\pm 0.42$& $196_{-63}^{+69}$ &1.71 &1987--1999 & \citet{Lasker08}\\
GTC &$r'$   & 0.623& 3631  & $17.61\pm 0.25$ & $328_{-67}^{+75}$& 1.77 & 2019-06-08--2019-08-07 & \citet{Torres19,Torres20}\\
USNO-B1.0 & $B$  & 0.442& 4067  & $17.94\pm 0.30$& $271_{-65}^{+86}$ &2.37 &1949--1965 & \citet{Monet03}\\
GSC2.3.2 & $B$  & 0.442& 4067  & $18.87\pm 0.41$& $115_{-36}^{+53}$ &2.37 &1987--2000 & \citet{Lasker08}\\
\hline
\end{tabular}
\tablecomments{See \citet{Straizys92} for the values of $F_0$. The magnitude for GTC gives the mid point of the observed range, and the uncertainty gives that range. }
\label{log}
\end{table*}

{\setlength{\tabcolsep}{3.5pt}
\begin{table*}\centering
\caption{The average UV--optical spectrum of \source observed on 2020-07-20--23
}
\vskip -0.4cm                               
\begin{tabular}{lccccccccc}
\hline
$\nu\,[10^{14}$\,Hz] & $16.0\pm 2.7$ & $13.5\pm 1.5$ & $11.7\pm 1.6$ & $8.76\pm 0.99$ & $6.82\pm 0.34$ & $5.51\pm 0.33$ & $4.87\pm 0.45$ & $4.01\pm 0.32$ & $3.36\pm 0.14$\\
$F_{\nu,{\rm obs}}\,[\mu$Jy] & $ 9.99\pm 1.17$&$ 14.0\pm 2.0$&$ 31.2\pm 3.3$&$ 59.5\pm 5.9$& $87.9\pm 11.3$& $148\pm 4$&$238\pm 15$& $303\pm 6$ &$415\pm 10$ \\
$10^{A_\lambda/2.5}$&5.39&7.65&4.28& 2.92& 2.27& 1.84 & 1.62 & 1.34 & 1.12 \\
\hline
\end{tabular}
\tablecomments{The measurements from 16.0 to $8.76\times 10^{14}$\,Hz are from the \swift Ultraviolet/Optical Telescope, and those from 6.82 to $3.36\times 10^{14}$\,Hz are from the Liverpool Telescope, La Palma \citep{Poutanen22}. The first row gives the middle frequencies and the band half-widths. }
\label{Juri}
\end{table*}}

\begin{figure}
\centerline{\includegraphics[width=\columnwidth]{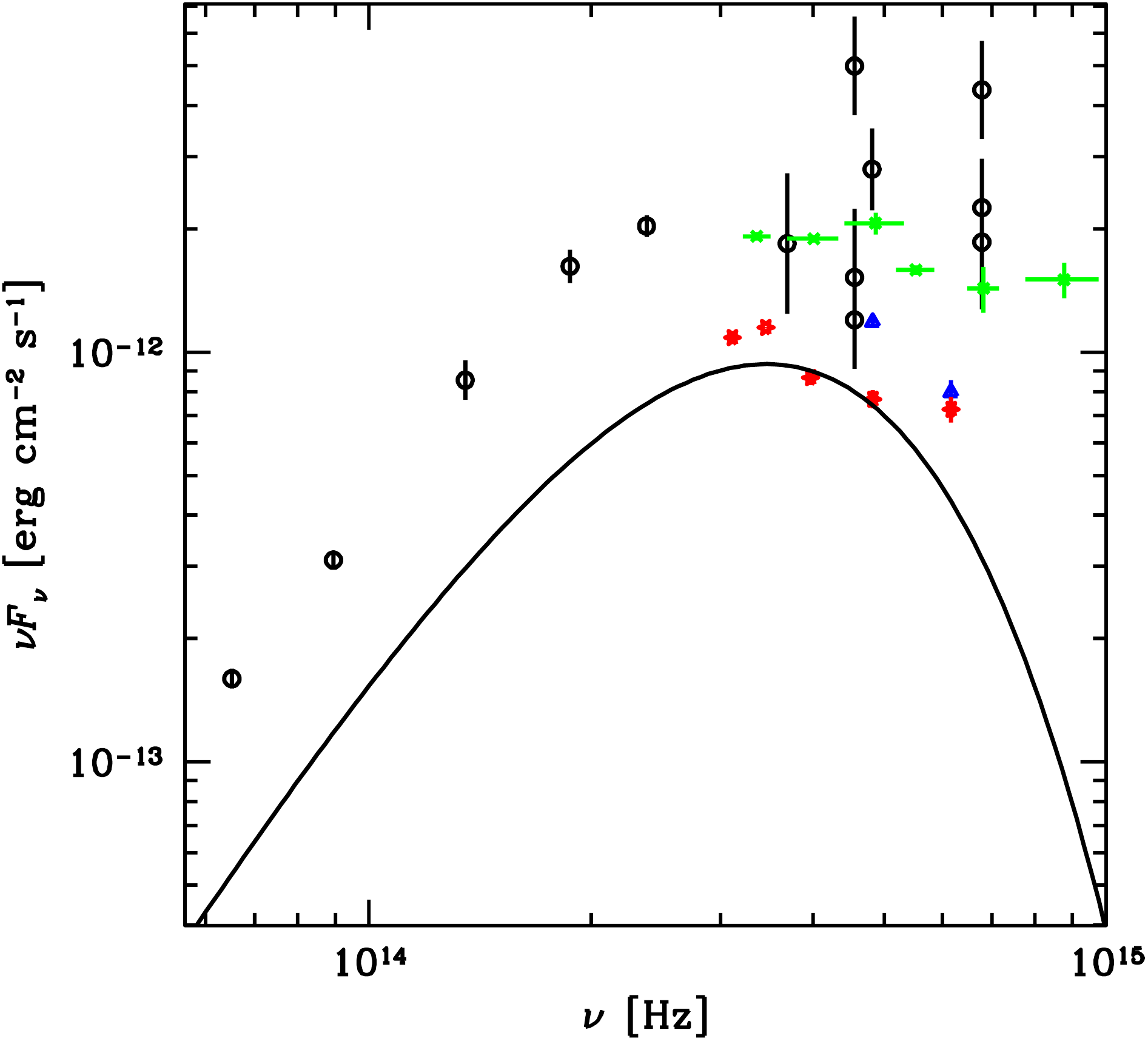}}
\caption{The extinction-corrected spectra of the binary during its quiescent states assuming $E(B$--$V)=0.23$. The red and blue symbols give the minimum fluxes measured by PS1 and ZTF, respectively; see Table \ref{pan}. The black error bars give the fluxes from the observations listed in Table \ref{log}. The green error bars show the average spectrum obtained by \citet{Poutanen22}, see Table \ref{Juri}. The black solid curve show the blackbody with $R_2=1.11 \rsun$, $D=3$\,kpc and $T=4230$\,K, whose $i$-band flux is at the upper limit of the PS1 measurement, see Equation (\ref{bb}) below. The amplitude of the flux variability due to ellipsoidal variability is $\lesssim \pm 10\%$ only.
}
 \label{spectrum}
 \end{figure}

\subsection{Photometric properties}
\label{photo}

We have compiled available ultraviolet to mid-infrared measurements of the binary during its quiescence from literature and public databases. These are listed in Tables \ref{pan} to \ref{Juri}, which also provide references to the photometric databases that were explored. In order to constrain the spectral distribution of the donor star, the collected fluxes need to be corrected for interstellar extinction. The total reddening through the Galaxy in the direction of the source has been estimated by \citet{Schlafly11} as $E(B$--$V)=0.197$ and as $E(B$--$V)=0.23$ by \citet{Schlegel98}. Since the source is located $\approx$0.5\,kpc above the Galactic disk (see Section \ref{metal}), the extinction toward it is likely to equal the total one in the direction of the target. We can also use the relationship between the  extinction in the $V$-band ($A_V$) and the H column density ($N_{\rm H}$) estimated from fitting X-ray data \citep{Guver09}, $N_{\rm H}  = (2.21 \pm 0.09) \times 10^{21} A_V\,{\rm cm}^{-2}$. Here $A_V$ is given by $A_V=3.1 E(B-V)$. The values of $N_{\rm H}$ obtained from fitting \xmm \citep{Kajava19}, NICER \citep{Bharali19} and the X-Ray Telescope \citep{Burrows00} onboard of \swift \citep{Wang21} data for \source are $1.4^{+0.3}_{-0.3}\times 10^{21}$,  $1.6^{+0.3}_{-0.2}\times 10^{21}$, $1.73^{+0.10}_{-0.07}\times 10^{21}$\,cm$^{-2}$, respectively. These values correspond (including the uncertainty of the conversion) to the range of $E(B$--$V)\approx 0.20$--0.26. To calculate the extinction at other wavelengths, $A_\lambda$, we used $A_V$ and the fitting formulae of \citet{Cardelli89}. Then, the observed magnitudes, $m_\lambda$, can be transformed into the the observed, $F_{\nu,{\rm obs}}$, and unabsorbed, $F_\nu$, fluxes, as
\begin{align}
F_{\nu,{\rm obs}}&=F_{0} 10^{-m_\lambda/2.5},\\
F_{\nu}&=F_{\nu,{\rm obs}} 10^{A_\lambda/2.5},
\label{fluxes}
\end{align}
where $F_{0}$ is the flux corresponding to a zero magnitude. 

\begin{figure}
\centerline{\includegraphics[width=\columnwidth]{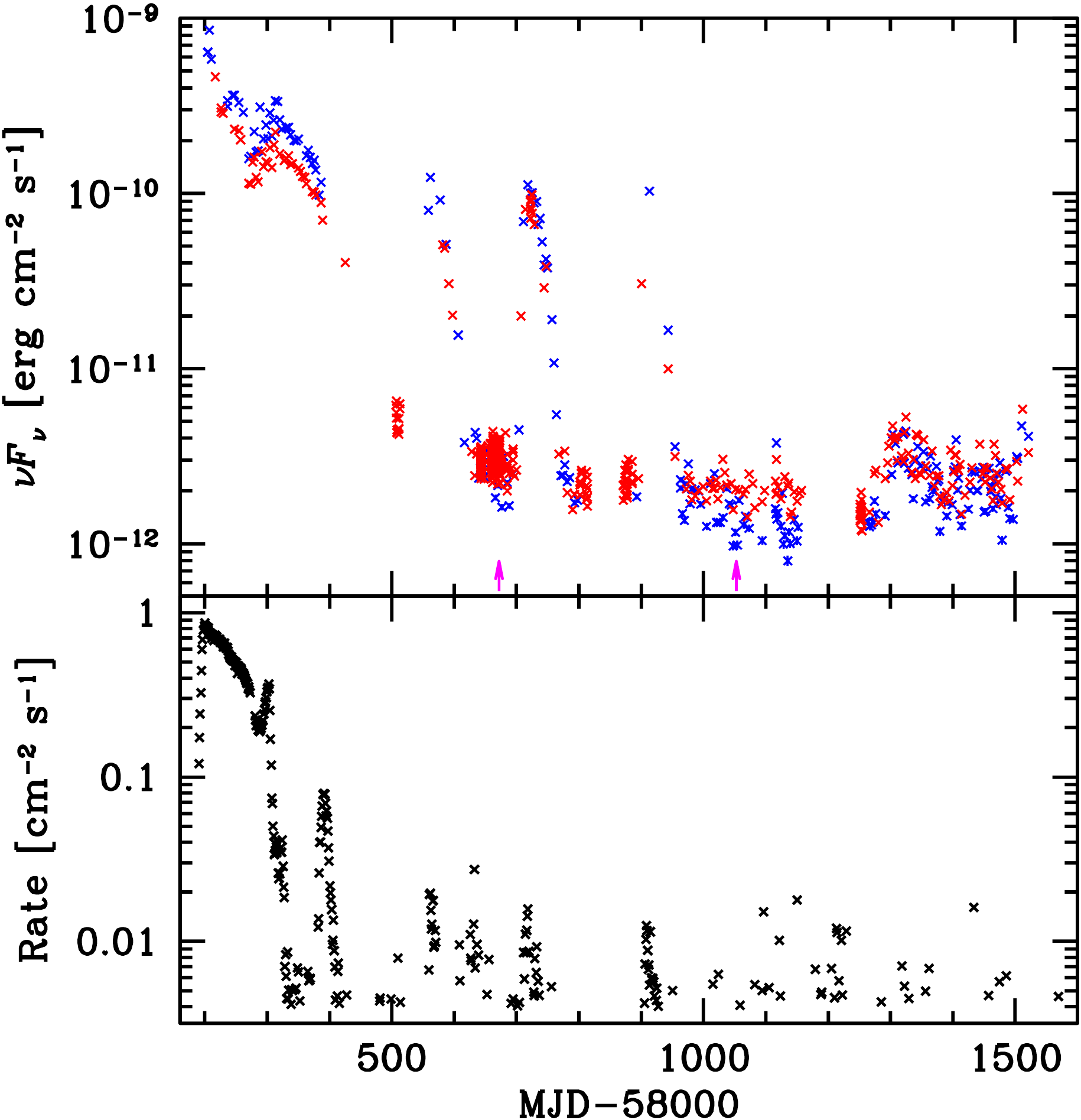}}
\caption{Top: the extinction-corrected [assuming $E(B$--$V)=0.23$] light curves of the binary from the ZTF. The red and blue symbols correspond to the $r$ and $g$ filters, respectively. The two magenta arrows indicate the middle times of the observations of \citet{Torres19,Torres20} and of \citet{Poutanen22}. Bottom: the 15--50\,keV daily-averaged count rate from the \swift BAT. This rate is shown only for $>$0.004\,cm$^{-2}$\,s$^{-1}$, at which boundary a typical fractional error is $\lesssim$0.3. We see that major reflares of the source were present until around MJD 58900, then followed by weaker activity in both optical and X-rays. 
}
 \label{ztf}
 \end{figure}

We first consider spectral data from Pan-STARSS1\footnote{\url{https://catalogs.mast.stsci.edu/panstarrs/}} Data Release 2 (PS1 DR2), which give measurements in the $y$, $z$, $i$, $r$ and $g$ bands for years previous to the outburst. We have searched for the overall minimum fluxes in each photometric band, using the {\tt psfFlux} and {\tt psfFluxErr} values in the Detection Table provided in PS1 DR2. Those fluxes were obtained using an automated pipeline; thus single detections at the faint end may be not real, but arise from a systematic noise \citep{Chambers16}, and be indistinguishable from real detections. In order to avoid them, we have applied a stringent selection criterion. Namely, we used {\it only\/} double detections from subsequent exposures separated by $\lesssim$0.5\,h and with the fluxes differing by $<$25\%. The results are given in Table \ref{pan} and plotted in red in Figure \ref{spectrum} after correcting for the extinction assuming $E(B$--$V)=0.23$. For comparison, Table \ref{pan} also gives the average fluxes from the DR2 stacked images, which are significantly higher than the minimum fluxes in the $i$, $r$ and $g$ bands.

Next, we have obtained the $r$ and $g$ light curves observed by the Zwicky Transient Facility\footnote{\url{https://irsa.ipac.caltech.edu/docs/program_interface/ztf_lightcurve_api.html}} (ZTF; \citealt{ZTF}), covering the discovery outburst and subsequent decay up to 2021 August. They are shown in Figure \ref{ztf}. We see that several mini-outbursts or reflares occurred after the end of the main outburst event, with the last major reflare in 2020 April. For comparison, we also show the hard X-ray light curve\footnote{\url{https://swift.gsfc.nasa.gov/results/transients/weak/MAXIJ1820p070/}} from the Burst Alert Facility (\citealt{BAT}; BAT) onboard of \swift. We see that the optical reflares have been accompanied by X-ray activity. From the ZTF light curves, we have obtained the weakest fluxes measured in each filter, which are given in Table \ref{pan} and are plotted in blue symbols in Figure \ref{spectrum}.  

Then, we have compiled some other available optical and infrared measurements of \source during quiescence. We list them in Table \ref{log} and plot in Figure \ref{spectrum} with black error bars. Table \ref{log} also gives the values of $F_{0}$. Most of the fluxes are higher than the minima from PS1. In particular, we analyzed the observations made during the states near quiescence in the $r$ band (of the SDSS photometric system) reported in \citet{Torres19,Torres20}. After calibrating the photometry of \source against the PSF mean magnitude of a non-variable star in the field, the range of those magnitudes is $r'\approx 17.87\pm 0.02$--$17.35\pm 0.02$, and its midpoint is $r'\approx 17.6$. Finally, we plot the average optical and UV fluxes obtained by \citet{Poutanen22}, see Table \ref{Juri}. The arrows in Figure \ref{ztf} indicate the middle times of the observations of \citet{Torres19,Torres20} and of \citet{Poutanen22}. 

\begin{figure}
\centerline{\includegraphics[width=\columnwidth]{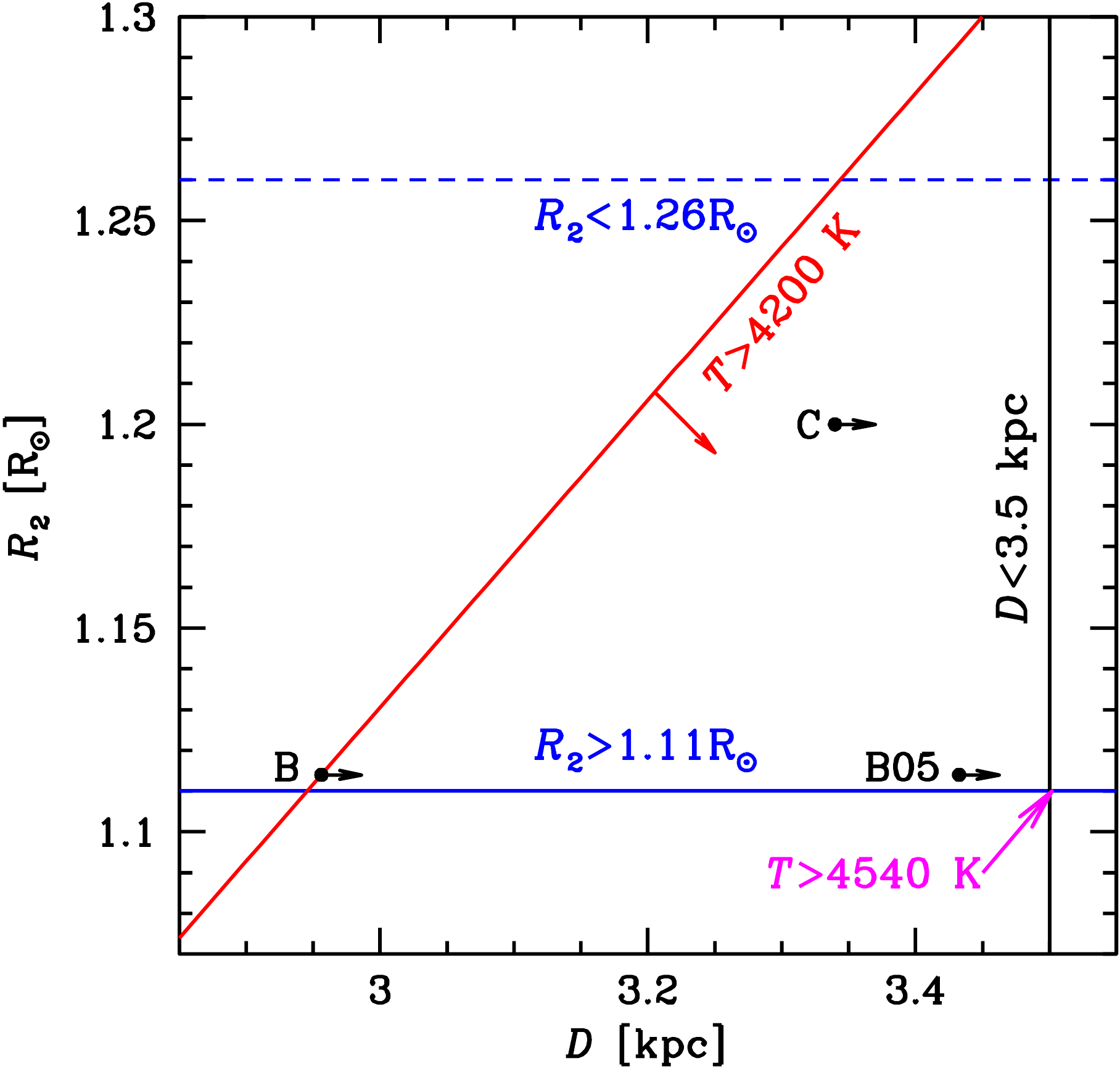}}
\caption{The region in the $D$--$R_2$ space allowed by the constraints on $R_2$ from the binary parameters, on $D$ from the \gaia parallax, and on $T$ from the blackbody [for $E(B$--$V)=0.23$] and spectral limits, see Equations (\ref{bb}--\ref{temp}). The horizontal blue solid and dashed lines correspond to the lower and upper limit on $R_2$ (Section \ref{binary}), respectively; the vertical black solid line gives the limit of $D\leq 3.5$\,kpc, and the diagonal red solid line corresponds to the limiting $D/R_2$ from the PS1 $i$-band flux and $T> 4200$\,K at $A=1$. If an accretion disk still contributes to that emission, $A>1$, and the red line moves in the direction of the red arrow becoming steeper. The allowed region is within the lower triangle delineated by the three solid lines. The limit on $T$ increases diagonally perpendicular to the red line from 4200\,K on it to 4540\,K at the lower right corner of the allowed space. Three evolutionary models of Section \ref{evolution}, B, B05 and C, satisfy these constraints, and are shown by the black solid circles with arrows (moving horizontally for $A>1$).
}
 \label{r2_d}
 \end{figure}

The PS1 DR2 minimum fluxes shown in Figure \ref{spectrum} constrain the stellar contribution, which we approximate as a blackbody, $F_\nu^*=\pi B_\nu(T)(R_2/D)^2$, where $B_\nu(T)$ is the blackbody intensity and $T$ is the effective temperature. The most restrictive measurement is that in the $i$ band, for which we take its upper limit (Table \ref{pan}) and assume $E(B$--$V)=0.23$. We have found that the blackbody relationship for this monochromatic flux can be locally approximated as a power law of $T\propto (D/R_2)^{0.45}$. The constraint, together with the temperature limit from our analysis of the average spectrum of \citet{Torres19,Torres20}, can be written as
\begin{align}
\frac{D/2.95\,{\rm kpc}}{R_2/1.11\rsun}&\gtrsim A\left(\frac{T}{4200\,{\rm K}}\right)^{2.22},\label{bb}\\
T&> 4200\,{\rm K},
\label{temp}
\end{align}
where Equation (\ref{bb}) is scaled to the lower limit on the stellar radius, see Section \ref{binary}. Here the constant $A$ accounts for the accretion disk being expected to be present at some level during the entire quiescence (see, e.g., the models of \citealt{Dubus01}), in which case $A>1$. On the other hand, the flux measurements may be affected by ellipsoidal variability, whose full amplitude we estimate as $<$0.2\,mag, i.e., less than $\pm 10\%$ with respect the average flux. This estimate is based on the ellipsoidal variability of the BH LMXB A0620--00, which has the mass ratio $q=0.067 \pm 0.01$ \citep{Marsh94} and $66\degr.5 < i < 73\degr.5$ \citep{Haswell93}, consistent within the uncertainties with those of \source. The full amplitude of the ellipsoidal variability of A0620--00 was measured by \citet{Marsh94} as 0.13\,mag. They also performed detailed simulations of this effect for a range of the inclinations, and their table 4 gives the theoretical range of 0.15--0.18\,mag at $i=70\degr$. Given the relative weakness of this effect, we neglect it. We also set $A=1$. The fiducial distance of 3\,kpc and the minimum $R_2$ correspond to $T\leq 4230$\,K. We plot a blackbody spectrum at $T= 4230$\,K in the solid black curve in Figure \ref{spectrum}. This spectrum gives a $\approx$26\% contribution [at $E(B$--$V)=0.23$] to the average flux in the $r$ band of \citet{Torres19,Torres20}. This is moderately higher than the stellar contribution of 16--21\% estimated by them. 

Equations (\ref{bb}--\ref{temp}) for $R_2\gtrsim 1.11\rsun$ imply a $D\gtrsim 2.95$\,kpc, in agreement with all the constraints on $D$ listed in Section \ref{binary}. On the other hand, these equations do not constrain $D$ from above. We take here $D\lesssim 3.5$\,kpc from the \gaia parallax as a tentative conservative upper limit. While the limit of \citet{Wood21} is stringent at face value, it relies on the two ejecta being exactly symmetric, which may not be the case, and 3.5\,kpc is within the $2\sigma$ uncertainty of the radio parallax, see Section \ref{binary}. We plot the resulting joint constraints on $D$ and $R_2$ in Figure \ref{r2_d}. At $D=3.5$\,kpc and $R_2=1.11\rsun$, we have the maximum $T\approx 4540$\,K, corresponding to a K5 stellar type. This satisfies the original constraint of \citet{Torres20}. We note that the above constraints favor low donor radii if $D\lesssim 3$\,kpc. At $D\leq 3$\,kpc and $T=4200$\,K, $R_2\leq 1.13\rsun$. Note that the above constraints can still be relaxed if there are some systematic error on the minimum PS1 fluxes, present in spite of our strict selection procedure.  

\subsection{Metallicity}
\label{metal}

\source appears to be a relatively old system and thus may have come from an environment with a subsolar metallicity. Its position in the Cartesian coordinates ($X,\,Y,\,Z$) centered on the Sun and the velocities calculated from the coordinated proper motion, ${\rm pmRA} = -3.093 \pm  0.091$\,mas/yr  and ${\rm pmDec} =-6.283 \pm 0.094$\,mas/yr (\gaia EDR3, \citealt{GaiaDR3}), the systemic velocity of $\gamma = -22$\,km/s \citep{Torres19} and the adopted distance of $D=3$\,kpc, are
\begin{align}
&X=2393\,{\rm pc},\, Y=1730\,{\rm pc},\, Z=529\,{\rm pc},\label{position}\\
&U=40.85\,{\rm km/s},V=-93.40\,{\rm km/s},\,W=-4.19\,{\rm km/s},\nonumber
\end{align}
where $U,\,V,\,W$ are velocities along the same axes, with $X,\,U$ positive towards the Galactic center, $Y,\,V$ positive in the direction of the Sun motion around the Galaxy, and $Z,\,W$ positive out of the plane of the Galaxy. \source is thus located in the thick Galactic disk as indicated by its distance from the Galactic plane, $Z=529$\,kpc, and a high peculiar velocity, $V_{\rm pec}=94$\,km/s, moving almost parallel to the Galactic plane. The Galactic $U,\,V,\,W$ velocities also locate it among the extended thick-disk population objects in Toomre diagram (e.g.\ fig. 1 in \citealt{Feltzing03}). The Galactocentric distance of \source is 5.9\,kpc, which locates it in the inner Galactic disk, where the  average value of the metallicity is $\langle [{\rm Fe/H}]\rangle =-0.55 \pm 0.17$ (e.g.\ \citealt{Bensby11}). The subsolar metallicity of the donor is further supported by relatively strong \ion{Ca}{1} and \ion{Ti}{1} absorption lines compared to \ion{Fe}{1} lines in the spectrum presented in fig.\ 3 of \citep{Torres19}, similar to other thick disk objects (e.g.\ \citealt{Feltzing03}).

We also note that high peculiar velocities of BH LMXBs are often attributed to a BH natal kick velocity, see, e.g. \citet{Atri19}.  That study included \source and three other systems with no known (at that time) systemic velocities and accurate distances, using instead a hypothetical kick velocity distribution modelled by Monte Carlo simulations. Based on them, they tentatively preferred the natal kick origin of the peculiar velocity in \source. Still, they also note that most LMXBs are relatively old systems, and thus may have come from low metallicity environments like globular clusters.

On the other hand, metallicities higher than solar were found in a number of donor stars in LMXBs (see review in \citealt{Casares17}). Still, this does not appear to be universal. One example is the binary V934 Her/4U 1700+24 consisting of a red giant accreting on a neutron star. Detailed photospheric abundance analysis for this system resulted in a slightly subsolar metallicity, ${[\rm Fe/H}] \approx -0.5$ and $[\alpha/{\rm Fe}]=+0.27$ (where $\alpha$ represents the average of the Mg, Si, and Ca elements), and the $[\alpha/{\rm Fe}]$ versus [Fe/H] relation close to the mean one without any notable peculiarity to this object \citep{Hinkle19}. Similar $\alpha$-element and Fe abundances were found in many symbiotic giants with white dwarf companions \citep{Galan16, Galan17}.

\subsection{Average mass transfer rate}
\label{fluence}

The bolometric flux of the source varied within only a factor of 2--3 during most of the 2018 discovery outburst, i.e., $\approx$200\,d, see, e.g., \citet{Shidatsu19}, \citet{Fabian20}, namely within $\approx(0.5$--$1.5)\times 10^{-7}$\,erg\,cm$^{-2}$\,s$^{-1}$. In order to estimate the total fluence of that outburst, we adopt the flux of $1.0\times 10^{-7}$\,erg\,cm$^{-2}$\,s$^{-1}$ during 200 d. Assuming $D=3$\,kpc and the accretion efficiency of $\epsilon=0.1$, we obtain the accreted mass of $M_{\rm accr} \approx 2\times 10^{25}$\,g. However, given the uncertainties on the fluence, distance and accretion efficiency, the uncertainty of this mass is probably by a factor of two.

After the 2018 discovery outburst, two historical outbursts in 1898 and 1934 were discovered by examining publicly available photographic plates \citep{Kojiguchi19}. Based on these data, the typical interval between outbursts of this source was estimated by those authors as 40 yr. This implies that the outburst within 1970--1980 was missed by observers of the period. However, intervals between outbursts of BH LMXBs are often unequal, and it is also possible that there was no outburst since 1934, giving the interval from the previous outburst of 84 yr. If we assume that the mass accreted onto the BH in 2018 equals to that transferred from the donor during the interval since the previous outburst (which theoretical assumption is clearly uncertain), the average mass transfer rate is $-\dot M_2\approx 1.2 \times 10^{-10}\msun$/yr, with an uncertainty, which we estimate to be by a factor of two. Alternatively, the estimate of $-\dot M_2$ would increase to $\approx 2.6 \times 10^{-10}\msun$/yr if the actual interval between outbursts was 40 yr.

\section{The evolutionary status}
\label{evolution}

We use the Warsaw stellar evolution code of \citet{Paczynski69,Paczynski70} described in \citet{Pamyatnykh98} and \citet{Ziolkowski05}. Our version of the code is calibrated to yield the H mass fraction of $X=0.74$, the metallicity of $Z=0.014$ (we use here the symbols $X$ and $Z$ in a different meaning than in Section \ref{metal}), and the mixing length parameter of 1.55 for the Sun at the solar age. The method of the evolutionary calculations in this work follow those used in \citet{zz17,zz18} and \citet{Zdziarski19}. We refer the reader to \citet{Zdziarski19} for details of the method.

\begin{figure}
\centerline{\includegraphics[width=\columnwidth]{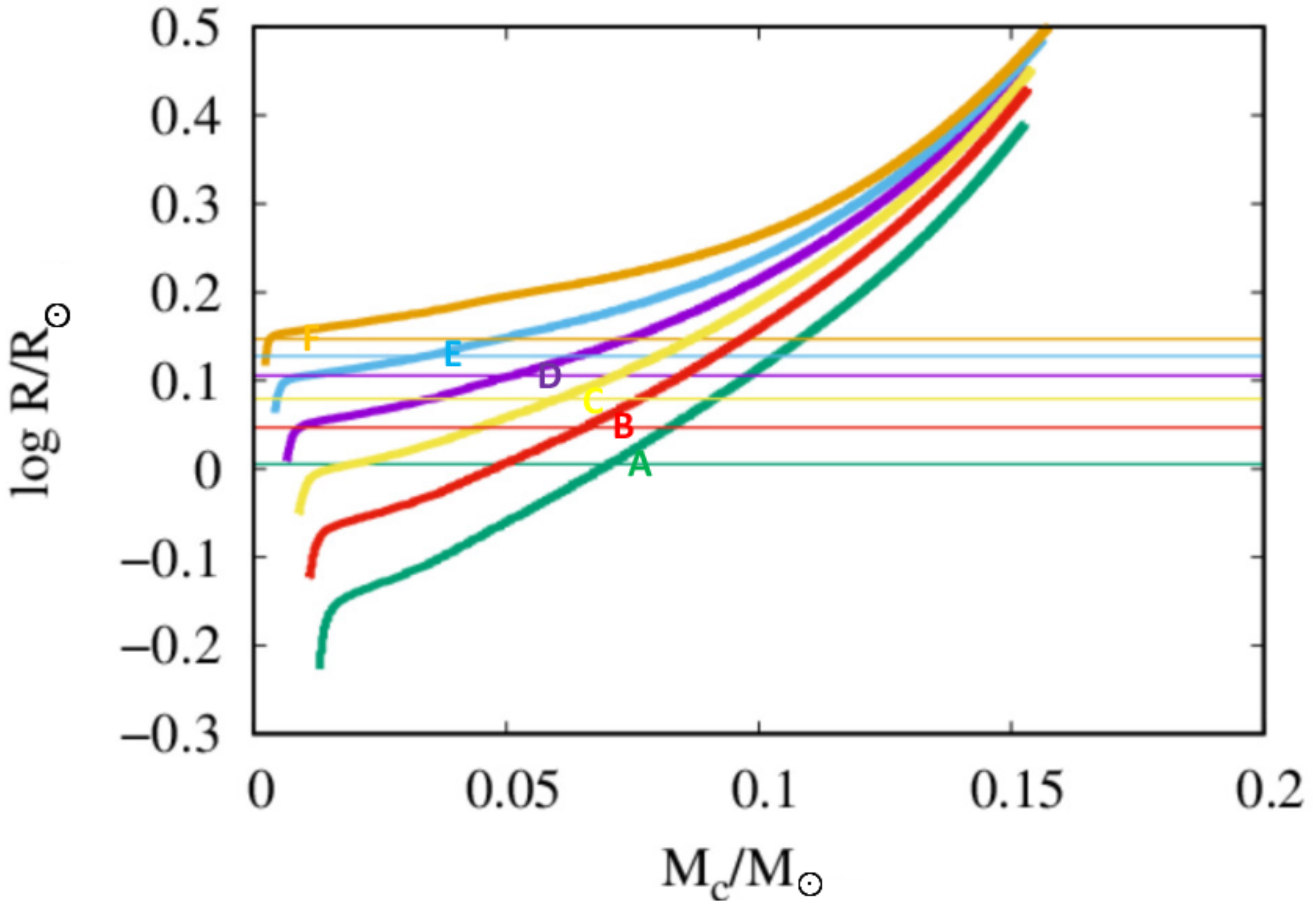}}
\caption{The radii of partially stripped, moderately evolved stars with solar metallicity vs.\ the mass of the He core for total donor masses of $M_2=0.3$, 0.4, 0.5, 0.6, 0.7 and $0.8 \msun$. They are shown (from bottom to top) by the green, red, yellow, violet, blue and orange thick solid curves, respectively. The evolution proceeds from left to right at the constant $M_2$. The horizontal dashed lines with the corresponding colors (from bottom to top) show the radii of the Roche lobe around the donor for the above values of $M_2$ (almost independent of $M_1$). The crossings of the corresponding evolutionary tracks and horizontal lines determine the positions of the models.
}
 \label{evolve}
 \end{figure}

\begin{table}
\begin{center}
\caption{The results of the evolutionary calculations for the donor
}
\label{models}
\begin{tabular}{lcccccc}
\hline Model &
$\displaystyle{\frac{M_2}{\msun}}$ &
$\displaystyle{\frac{R_2}{\rsun}}$ &
$\displaystyle{\frac{L_2}{\lsun}}$ &
$T\,[{\rm K}]$ &
$\displaystyle{\frac{M_{\rm c}}{\msun}}$ &
$\displaystyle{-\dot M_{2,10}}$\\
\hline
A  &0.3 &1.012 &0.278 &4169 & 0.070& 0.43 \\
A2 &0.3 &1.012 &0.219 &3924 & 0.290& 0.155\\
{\bf B05}&0.4 &1.114 &0.452 &4492 & 0.050& 1.0 \\
{\bf B}  &0.4 &1.114 &0.347 &4200 & 0.065& 0.68 \\
B2 &0.4 &1.114 &0.273 &3948 & 0.084& 0.27 \\
C05&0.5 &1.200 &0.587 &4621 & 0.042& 1.3 \\
{\bf C}  &0.5 &1.200 &0.437 &4289 & 0.060& 0.91 \\
D  &0.6 &1.275 &0.560 &4430 & 0.051& 1.1  \\
D2 &0.6 &1.275 &0.414 &4102 & 0.130& 0.49 \\
E  &0.7 &1.343 &0.749 &4643 & 0.034& 1.25 \\
F  &0.8 &1.404 &1.019 &4903 & 0.003& 1.55 \\
\hline
\end{tabular}
\tablecomments{$M_{\rm c}$ is the mass of the He core, and $-\dot M_{2,10}$ is the stellar mass loss rate in units of $10^{-10} \msun\,{\rm yr}^{-1}$. The BH mass of $M_1=7\msun$ was assumed. The models denoted by a single letter have the solar metallicity $Z=0.014$, while models A2, B2 and D2 have $Z=0.028$, and B05 and C05 have $Z=0.007$. Model B is at the minimum temperature allowed by the GTC spectrum and corresponds to $D\gtrsim 2.9$\,pc. Model B05 has the temperature approximately best-fitting the spectrum, but is requires $D\gtrsim 3.4$\,kpc. Model C has an intermediate temperature and $D\gtrsim 3.3$\,kpc. These models are marked in Figure \ref{r2_d}.
}
\end{center}
\end{table}

We have first followed the evolution in the core mass--stellar radius plane of partially stripped stars for total masses of $M_2=0.3$, 0.4, 0.5, 0.6, 0.7 and $0.8 \msun$ and at the solar metallicity. The results are shown in Figure \ref{evolve} and Table \ref{models}, where these models are denoted by letters A, B, C, D, E and F, respectively. In Figure \ref{evolve}, the evolution proceeds from left to right at the constant $M_2$. The evolutionary engine lies in the hydrogen burning shell, which adds newly synthesized helium to the core, increasing its mass. The horizontal thin lines with the corresponding colors (from bottom to top) show the radii of the Roche lobe around the donor for the above values of $M_2$, and $M_1=7\msun$. The crossings of the corresponding evolutionary tracks and horizontal lines determine the positions of our donor models. The relatively small core masses of those models confirm that the donor is in an early subgiant phase, with the radius only a factor of about two larger than a main-sequence star of the same mass (as noticed in Section \ref{binary}). Given that the donor metallicity is relatively uncertain (Section \ref{metal}), we also calculated some models for a metallicity of half solar, $Z=0.007$ (B05 and C05), and twice solar, $Z=0.028$, (A2, B2 and D2), whose results are also given in Table \ref{models}. The seen anticorrelation of the metallicity and temperature in our evolutionary models is a well-known effect, resulting from an increase of the opacities with the metallicity, present also in main sequence stars. 

\begin{figure}[t!]
\centerline{\includegraphics[width=\columnwidth]{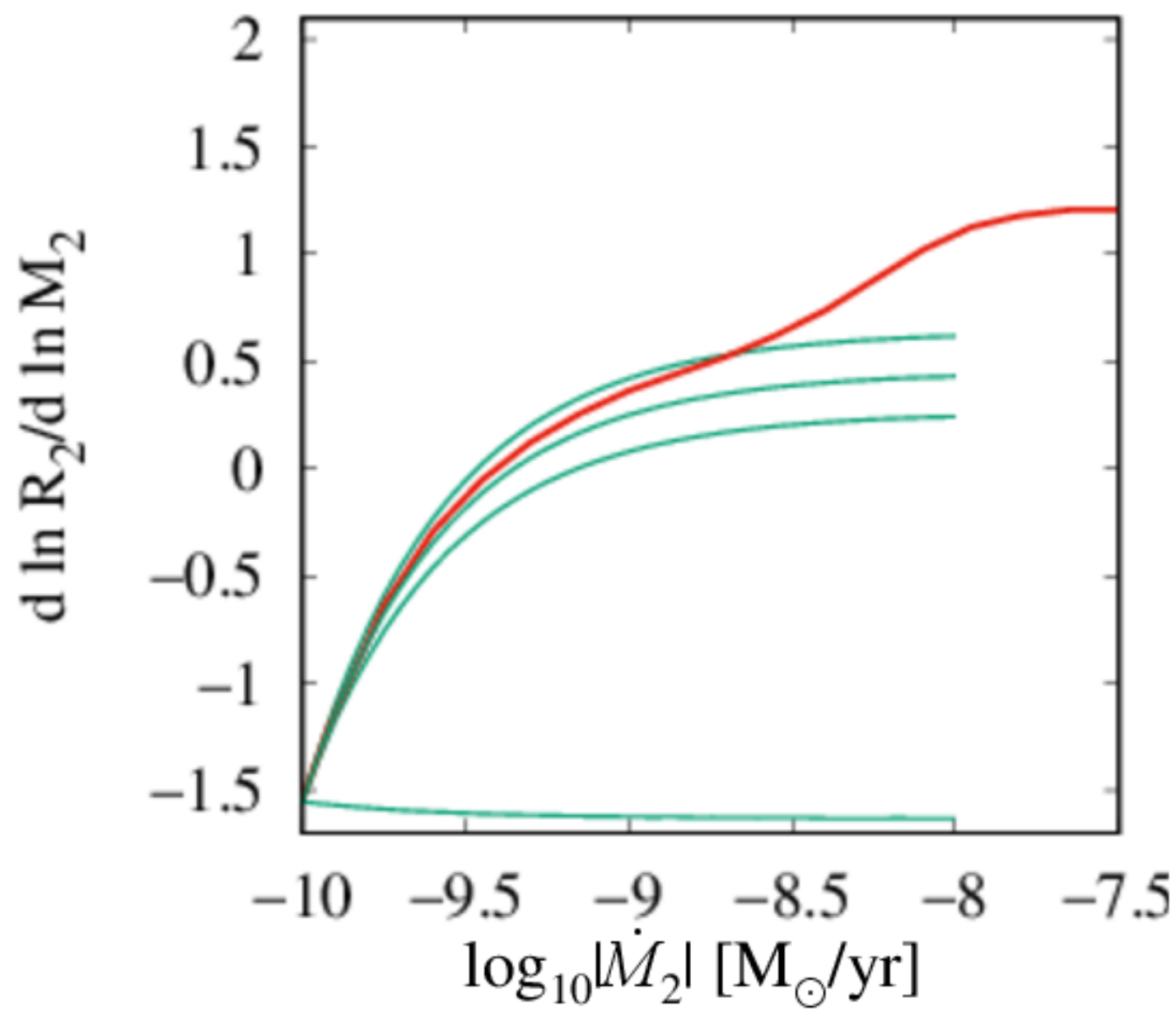}} 
\caption{The rates of the evolutionary change of the donor radius (red curve) and  that of its Roche lobe (green curves) as functions of the donor mass loss rate, $-\dot M_2$. The conservative solution corresponds to model B05, with the averaged accretion rate on the BH of $\langle\dot M_1\rangle= 1.0\times 10^{-10}\msun$\,yr$^{-1}$. The results of our evolutionary model include now the effect of mass outflow from the outer layers of the star. The green curves show ${\rm d}\ln  R_{\rm 2,L}/{\rm d}\ln M_2$ for $\alpha =0$, 1, 1.1, and 1.2, from bottom  to top. The values of $\beta$ along those curves are given by $\langle\dot M_1\rangle/|\dot  M_2|$, i.e., $\beta$ =1 at the beginnings of the green curves. The intersection of the red curve and the top green one gives a possible self-consistent and non-conservative model of \source, see Section \ref{evolution}.} 
\label{derivative}
\end{figure}

The assumed BH mass, $M_1$, influences only one model parameter, namely the rate of the mass transfer, $-\dot M_2$. However, this effect is rather weak. For models A and F, changing $M_1$ to $3.17\msun$ and $11.1\msun$, respectively (to keep the mass ratio equal 0.072), would change $-\dot M_2$ from 4.3 to $4.4\times 10^{-11} \msun$/yr, and from 1.55 to $1.50\times 10^{-10} \msun$/yr, respectively. 

We compare now the models of Table \ref{models} with the Monte Carlo constraints obtained in Section \ref{binary} and other constraints obtained in Sections \ref{photo}, \ref{metal} and \ref{fluence}. As shown based on the fluxes measured before the outburst (Section \ref{photo}), an approximate maximal stellar blackbody contribution at $D=3$\,kpc has $T=4200$\,K and $R_2=1.13\rsun$. This closely corresponds to our model B, which support an approximate solar metallicity of the donor. If we still impose the spectral type to be within K5--K3 (which requires $D\sim 3.5$\,kpc), the temperature has to be larger, which approximately corresponds to our model B05 with $T=4492$\,K. This model has half-solar metallicity, which agrees with the finding of Section \ref{metal} that \source is likely to belong to extended Galactic thick disk population. On the other hand, it is in tension with the X-ray spectral fitting of \citet{Zdziarski21b}, which favor the metallicity to be at least solar. Our model C, with $M_2=0.5\msun$ and $R_2=1.2\rsun$, also satisfies our constraints, yielding $T=4289$\,K and $D\gtrsim 3.3$\,kpc. These three models are shown in the $R_2$--$D$ parameter space in Figure \ref{r2_d}. We see that the constraints shown on Figure \ref{r2_d} allow the metallicity to be only slightly above solar. On the other hand, our models with metallicity twice solar, A2, B2 and D2, have the temperatures of 3924\,K--4102\,K, which are below our strict lower limit of 4200\,K.

The calculated mass-loss rate for model B05 is $-\dot M_2\approx 1.0\times 10^{-10} \msun$/yr, which agrees well with that estimated in Section \ref{fluence} of $1.2\times 10^{-10} \msun$/yr with an uncertainty by a factor of two. For model B, $-\dot M_2$ is only slightly lower. 

The above agreement implies that the mass loss rate approximately equals the accretion rate averaged over the interval since the previous outburst, $\langle\dot M_1\rangle$, i.e., the mass transfer is conservative. However, this assumption is uncertain, with outflows and not full exhaustion of the disk during outburst being possible. Indeed, mass loss from a disk wind in \source was found by \citet{Munoz19}. This motivates us to consider also models with non-conservative mass transfer. We note that taking this effect into account does not change the model parameters of the donor. It only increases the model values of $-\dot M_2$. We consider here non-conservative mass transfer for model B05, following the approach of \citet{zz18}. Such transfer is described by two parameters, $\alpha$ and $\beta$. As in  \citet{zz18}, we define $\alpha$ as the specific angular momentum of the mass leaving the system in units of the specific angular momentum of the donor measured from the center of mass \citep{Verbunt93}, with $\alpha\lesssim 1$ (as discussed, e.g., in \citealt{zz18}). The parameter $\beta$ is defined as the fraction of the mass lost by the donor that is accreted onto the accretor \citep{Rappaport82}, i.e., $-\dot M_2=\langle\dot M_1\rangle/\beta$. 

We assume $\langle\dot M_1\rangle= 1.0\times 10^{-10}\msun$\,yr$^{-1}$. The results are shown in Fig.\ \ref{derivative}. We compare here the rates of the Roche-lobe change for $\alpha= 0$, 1, 1.1 and 1.2 with the rate implied by our evolutionary model as a function of $\beta=\langle\dot M_1\rangle/|\dot M_2|$. Apart from the conservative solution ($\beta=1$), a non-conservative solution is possible for $\beta\approx 0.05$ and $\alpha=1.2$. A caveat for it is its relatively large $\alpha$, requiring the outflow to occur far from the BH, around the disk tidal radius (see fig.\ 1 of \citealt{zz18}). Thus, the simplest and most likely solution remains the conservative case.

\section{Summary}
\label{summary}

We have used the 1-$\sigma$ deviations for the binary parameters of \source and a uniform distribution in $\cos i$ for the range of inclinations given in \citet{Torres20} to calculate the donor star mass and radius.  We obtain 68\% confidence values of $M_2= 0.49^{+0.10}_{-0.10}\msun$ and $R_2= 1.19^{+0.08}_{-0.08}\rsun$ (as well as $M_1= 6.75^{+0.64}_{-0.46}\msun$ for the BH mass).

We have compiled available optical and infrared measurements of \source in quiescence, which constrain the spectrum of the donor. The strongest constraint is given by the minimum fluxes from PS1 DR2 in the $i$ and $r$ bands, which gives the maximum allowed blackbody temperature of $T< 4230$\,K at the lower limit on the donor radius allowed by the binary parameters ($R_2=1.11\rsun$), $D\approx 3$\,kpc and $E(B$--$V)=0.23$. This $T$ is at our lower limit of the temperature compatible with the average GTC spectrum of $T>4200$\,K, and it is less than those implied by the original classification of the stellar type of the donor within the range of K3--5 \citep{Torres20}. Requiring the latter can be reconciled with our constraints if $D\sim 3.5$\,kpc, yielding $T\lesssim 4540$\,K, of a K5 stellar type. The distance of $D=3.5$\,kpc is allowed by the \gaia parallax, but it is larger than those obtained from the radio parallax and the moving ejecta method.

We then considered the source location in the Galaxy and peculiar velocity, which indicate \source belongs to the extended thick-disk population. That population is characterized by the metallicity less than solar. We also estimated the fluence during the 2018 outburst. Using the time elapsed since the previous detected outburst of the system, we evaluate the average mass accretion rate (equal to the average transfer rate for conservative accretion) of $-\dot M_2\sim 10^{-10}\msun$/yr.

We finally considered a range of stellar evolutionary models that may correspond to \source. Our model B (see Table \ref{models}) approximates the above constraints at the limiting $T=4200$\,K, the donor mass of $M_2=0.4\msun$ and the solar metallicity (which approximately agrees with the X-ray spectral fitting results of \citealt{Zdziarski21b}) and requires $D\geq 3$\,kpc. An alternative at $D\gtrsim 3.3$\,kpc is our model C with $T\approx 4300$\,K. Another possibility at $D\gtrsim 3.3$\,kpc is our model B05, with $T\approx 4500$\,K and the metallicity half solar, which, in turn, agrees with the location of the source in the Galaxy. Thus, we cannot unambiguously constrain the donor metallicity. Still, we find models with the metallicity twice solar to be ruled out.

We also considered evolutionary models with non-conservative accretion. However, we have found they require an unrealistically large specific angular momentum of the mass leaving the system. 

\section*{Acknowledgments}

We thank Juri Poutanen for providing us with their average optical-ultraviolet spectrum of \source, and the referee for valuable suggestions. We acknowledge support from the Polish National Science Centre under the grants 2015/18/A/ST9/00746, 2019/35/B/ST9/03944 and 2017/27/B/ST9/01940. AAZ acknowledges support from the International Space Science Institute (Bern). MAPT and JCV have  been supported by the Spanish MINECO under grant AYA2017-83216-P and PID2020-120323GB-I00. MAPT acknowledge support via a Ram\'on y Cajal Fellowship RYC-2015-17854.  

\bibliography{../../allbib}{}
\bibliographystyle{aasjournal}

\label{lastpage}
\end{document}